\documentclass[5p,twocolumn,sort&compress,final]{elsarticle}
\usepackage{graphicx,tabularx,epsf,subfigure,amsbsy,amssymb,amsmath,float}
\usepackage[latin1]{inputenc}
\usepackage{epstopdf}
\usepackage{algorithm}
\usepackage{algpseudocode}

\newcommand{\pluseq}{\mathrel{+}=}

\journal{Computer Physics Communications}

\begin{document}

\title{Efficient implementation of core-excitation Bethe-Salpeter equation calculations}

\author[kg1,kg2]{K.~Gilmore\corref{cor1}}
\ead{kgilmore@esrf.fr}

\author[NIST]{John Vinson}
\author[NIST]{E.L.~Shirley}

\author[LBNL]{D.~Prendergast}
\author[LBNL]{C.D.~Pemmaraju}

\author[UW]{J.J.~Kas}
\author[UW]{F.D.~Vila}
\author[UW]{J.J.~Rehr}

\address[kg1]{European Synchrotron Radiation Facility (ESRF), Grenoble 38000, France}
\address[kg2]{Jiangsu Key Laboratory for Carbon-Based Functional Materials \& Devices, Institute of Functional Nano \& Soft Materials (FUNSOM), Soochow University, Suzhou, Jiangsu 215123, PR China}
\address[NIST]{National Institute of Standards and Technology (NIST), Gaithersburg, Maryland 20899, USA}
\address[LBNL]{Lawrence Berkeley National Lab (LBL), Berkeley, California 94720, USA}
\address[UW]{Department of Physics, University of Washington, Seattle, Washington 98195, USA}

\cortext[cor1]{Corresponding author at: European Synchrotron Radiation Facility, 38043 Grenoble, France. Tel.: +33 476881727.}

\date{\today}

\begin{abstract}
We present an efficient implementation of the Bethe-Salpeter equation (BSE)
method for obtaining core-level spectra including x-ray absorption (XAS), x-ray 
emission (XES), and both resonant and non-resonant inelastic x-ray scattering spectra 
(N/RIXS).  Calculations are based on density functional theory (DFT) electronic structures 
generated either by {\sc abinit} or Quantum{\sc espresso}, both plane-wave basis, 
pseudopotential codes.  This electronic structure is improved through the inclusion of a 
{\it GW} self energy.  The projector augmented wave technique is used to evaluate transition 
matrix elements between core-level and band states.  Final two-particle scattering states 
are obtained with the NIST
core-level BSE solver (NBSE).  We have previously reported this implementation, which we
refer to as {\sc ocean} (Obtaining Core Excitations from {\em Ab initio} electronic
structure and NBSE)
[{\em Phys.~Rev.~B} {\bf 83}, 115106 (2011)].  Here, we present additional efficiencies
that enable us to evaluate spectra for systems ten times larger than previously possible;
containing up to a few thousand electrons.  These improvements include the implementation 
of optimal basis functions that reduce the cost of the initial DFT calculations, 
more complete parallelization of the screening calculation and of the action of the BSE 
Hamiltonian, and various memory reductions.  Scaling is demonstrated on supercells of 
SrTiO$_{3}$ and example spectra for the organic light emitting molecule 
Tris-(8-hydroxyquinoline)aluminum (Alq$_3$) are presented.  The ability to perform large-scale 
spectral calculations is particularly advantageous for investigating dilute or non-periodic 
systems such as doped materials, amorphous systems, or complex nano-structures.
\end{abstract}

\maketitle

\section{Introduction}

Core-level spectroscopies provide a quantitative element- and orbital-specific probe 
of the local chemical environment and the atomic and electronic structure of materials.  
For example, X-ray absorption (XAS) and emission (XES) spectra probe the unoccupied 
and occupied densities of states, respectively.  The near-edge absorption
region (XANES) is sensitive to oxidation state, spin configuration, crystal
field, and chemical bonding, whereas the extended region can be used to reconstruct 
the local coordination shells.  However, extracting this information requires a 
reliable interpretation of the measured spectra.  Often this is done by comparison 
with reference spectra, but such comparisons are at best qualitative; it is preferable 
to calculate spectra quantitatively and predictively.

An accurate description of core-level excitations must take into account both the highly localized nature of the core hole and the extended condensed system. The problem of predictive computational x-ray spectroscopy has been approached from many directions, but most can be divided by scale into two major categories. Atomic and cluster models have sought to include a more exact treatment of many-body effects by considering a small subsystem coupled loosely or parametrically to the larger system. At the other end, various single-particle theories are able to treat many hundreds of atoms by approximating the electron-electron and electron-hole interactions. The utility of calculated spectra hinges upon compromises between the ability to accurately model a given system and the ability to address systems large enough to be representative of experiment: defects, dopants, interfaces, etc. 

Within a first-principles approach, it is easiest to use the independent-particle 
approximation.  For x-ray absorption, the extended region of materials 
is very well reproduced by the real-space Green's function code {\sc feff}, which is 
widely used over extensive energy ranges \cite{rehr.compt-rend-phys}.  
However, the current implementation loses accuracy at the edge when non-spherical 
corrections to the potential are important.  To reproduce the near-edge structure, 
accurate independent quasi-particle models for deep-core XAS 
 can be constructed for 
$s$-levels \cite{rehr.fs-vs-bse}.  In this case, the x-ray absorption intensity is proportional to the 
unoccupied projected density of states in the presence of a screened core-hole, 
weighted by final-state transition matrix elements.
This approach has been used with different treatments of the core-hole including 
the ``Slater transition state" model of a half-occupied core state \cite{pettersson.hch} 
or the final state picture of a full core hole with \cite{prendergast.xch} or without 
\cite{pickard.fch} the corresponding excited electron.
Due to the simplicity of their implementation and modest computational cost, such 
core-hole schemes have been implemented in several standard DFT distributions 
\cite{abinit0, abinit1, abinit2, abinit3, espresso1, espresso2, wien2k} for near-edge spectra.  
However, these approaches often fail when the hole has non-zero orbital
angular momentum.  Extensions to DFT such as time-dependent DFT (TDDFT) have been problematic 
due to the lack of sufficiently accurate exchange-correlation functionals for core-excited 
states.  Although recent development of more accurate short-range exchange-corrected 
functionals \cite{besley.2009, besley.2010} has improved the viability of TDDFT for 
calculating core-level spectra more advances will be needed to make TDDFT a generally 
applicable approach.

%Unfortunately, extensions of DFT such as time-dependent DFT (TDDFT)
%are problematic due to the lack of accurate exchange-correlation functionals for 
%highly excited states.

%To obtain a more general prescription for calculating x-ray spectra
%a many body treatment is necessary. 
%In particular, calculations require an accurate treatment of deep core-hole and
%photoelectron states, including self-energy effects and the
%``excitonic" interaction between these states.  Multi-electron excitations
%can also be important in certain cases.
%Contrary to DFT-based methods that typically begin from the perspective of 
%the extended system, atomic and cluster models can more easily include many-electron 
%effects.  In particular, atomic multiplet theory \cite{degroot.book} 
%or configuration interaction \cite{ikeno.ci,Miedema.ci} and exact diagonalization methods 
%\cite{devereaux.xas} can accurately reproduce complicated L edges of transition 
%metal systems.  However, these techniques that take a local description typically
%ignore solid-state effects and are limited to small cluster models, although recent 
%progress at incorporating band structure within the multiplet method should be noted 
%\cite{haverkort,PhysRevB.83.155107,Josefsson2012}.

The aforementioned approaches are all inherently single-particle descriptions. 
For calculating x-ray spectra this can be problematic, not only when the core hole has non-zero angular 
momentum, but also when many-body or multi-electron excitations are important. 
Many-electron wavefunction-based methods constitute one obvious approach to this problem. 
Contrary to single-particle descriptions that typically begin from the perspective of the extended system, 
atomic and cluster models start with the goal of completely describing the local problem. 
In particular, atomic multiplet theory \cite{degroot.book} 
or configuration interaction \cite{ikeno.ci,Miedema.ci}  
and exact diagonalization methods \cite{devereaux.xas} can accurately reproduce complicated L edges 
of transition metal systems. However, these techniques that take a local description 
typically ignore solid-state effects and are limited to small cluster models, 
although recent progress at incorporating band structure within the multiplet model should be noted 
\cite{haverkort,PhysRevB.83.155107,Josefsson2012}.

Further improvements in DFT-based approaches to calculating spectra require 
a two-particle picture including particle-hole interactions, particle and 
hole self-energies, and full-potential electronic structure,
within the context of many-body perturbation theory.  Specifically, 
this involves solving the Bethe-Salpeter equation (BSE), i.e., a
particle-hole Green's function.  The Bethe-Salpeter equation description of  
absorption includes single-particle terms that describe the
quasi-particle energies of 
the core hole and the excited photoelectron, together with the interaction
between them.  To leading order
the interaction consists of two terms: the Coulomb interaction, which 
includes adiabatic screening of the core hole, and
an unscreened exchange term.  Even this two-particle description of the
many-body final state is already a significant improvement 
for L edges \cite{shirley.bse-xas}. 
For example, when considering the L$_{2,3}$ edges of the 
transition metals, the independent-particle approximation predicts a
2:1 branching ratio 
between the intensities of the L$_{3}$ and L$_{2}$ edges, which is in
contrast to experiments which exhibit branching ratios ranging from 0.7:1
for Ti to beyond 2:1 for Co and Ni \cite{leapman.branching,fink.branching,
haverkort.branching, vanderlaan.branching, 
prince.branching, chen.branching-fe-co, chen.branching-ni}. 
The BSE largely resolves this discrepancy, yielding branching ratios
in reasonable agreement with experiment \cite{vinson.metals}. However,
simpler approaches such as TDDFT can also account for these 
corrections \cite{EbertTDDFT,Ankudinov.2003,Ankudinov.2005}.

BSE solvers have been implemented in a few core-level codes 
to date \cite{draxl.ae-bse, kruger.bse,blaha.bse} -- as well as some valence 
level codes \cite{lawler.ai2nbse, sottile.exc,BerkeleyGW,yambo} -- but their utility has been
limited to a specialist community.  In part, this
is due to significantly increased computational cost.  
{\sc feff} and DFT-based core-hole approaches require little more effort
than standard DFT calculations \cite{core-hole}, and calculations 
on systems of hundreds of atoms have become routine.
BSE calculations are considerably more intensive and have heretofore 
typically been limited to a few tens of atoms.  This added cost 
is largely associated with including {\it GW} self-energy corrections to the electronic structure, 
obtaining the screening response to the core hole, and acting with the Bethe-Salpeter Hamiltonian 
on the electron-hole wavefunction to obtain the excitation spectrum.  However, 
given the significantly improved accuracy of the BSE method it is desirable to
make this a more widely used technique.  This necessitates improving its ease of
use and reducing the computational cost.  Toward this second 
objective we report herein several efficiency improvements 
to our existing BSE code \cite{vinson.ocean} that now allow BSE calculations 
on systems of hundreds of atoms and significantly reduce the time required for 
previously viable smaller systems.

The most time-consuming steps of a BSE calculation are (1) obtaining the ground-state electronic 
structure, (2) correcting the quasiparticle energies by adding a ({\it GW}) self-energy, (3) evaluating 
the screening response to the core-hole, and (4) determining the excitation spectrum of the BSE Hamiltonian.
Our BSE calculations build on self-consistent field (SCF)
DFT calculations of the ground-state charge density and the
accompanying Kohn-Sham potential. 
We then use non-self-consistent field (NSCF) calculations,
i.e., direct calculations that solve the one-electron Schr\"{o}dinger
equation in the already computed Kohn-Sham potential,
to obtain all desired occupied and unoccupied Bloch states. 
To alleviate the burden of {\bf k}-space sampling during the NSCF calculation, and 
to reduce the plane-wave basis, we implement a {\bf k}-space interpolation scheme that solves a 
$k$-dependent Hamiltonian over a reduced set of optimal basis functions \cite{shirley.obf, 
prendergast.obf}.  This is described in section $3.1$.  In most cases, rather than evaluating the 
{\it GW} self-energy in the typical random-phase approximation ($G^{0}W^{\text{RPA}}$), we instead use a much 
more computationally efficient approximation based on a multi-pole model for the loss function.  
This has been previously described in detail \cite{kas.mpse}, and we will not discuss it further here.  To 
reduce the time required to calculate the screening response to the core-hole we take advantage of 
the fact that this screening is highly localized around the excited atom and partition space accordingly.  
The electronic response is evaluated locally and a model dielectric response proves adequate for 
the rest of space \cite{shirley.screening}.  Here we reduce the time needed to evaluate the screening response 
and action of the BSE Hamiltonian by parallelizing these portions of the code.  This is discussed in 
section \ref{bseHam} for the BSE Hamiltonian and section \ref{screen} for the screening.  In section 4 we demonstrate 
the effectiveness of these improvements through XAS calculations on a series of supercells of SrTiO$_3$.
Section 4.1 characterizes the time-scaling of the code with respect to system size.  Section 4.2 
evaluates the savings realized by employing the optimal basis functions.  The efficacy of parallelization 
is reported in section 4.3 for the action of the BSE Hamiltonian and in section 4.4 for the screening 
response.  Example XAS and XES spectra of the commonly studied organic molecule 
Tris-(8-hydroxyquinoline)aluminum (Alq$_3$) are presented in section 4.5.  We end with a summary of the 
capabilities of {\sc ocean} and some general comments on its applicability.

\section{Formalism}

The theoretical description of the absorption of a photon by a material may be 
expressed in terms of the loss function $\sigma({\bf q},\omega) = -{\rm Im} \, \epsilon^{-1}({\bf q}, \omega)$.  
The dielectric response function $\epsilon$  depends on the 
photon energy $\omega$ and the momentum transfer ${\bf q}$.  A formal many-body expression 
for the loss function may be given as
~
\begin{equation}  \label{loss.fcn}
\sigma({\bf q},\omega) \propto -\frac{1}{\pi} {\rm Im} \langle 0 | \hat{O}^{+} G(\omega) \hat{O}  | 0 \rangle ,
\end{equation}

\noindent where $| 0 \rangle$ is the many-body ground-state wavefunction, the operator $\hat{O}$ 
describes the interaction between the photon field and the system, and $G(\omega)$ is the Green's 
function for the many-body excited-state.  The form used for the operator $\hat{O}$ depends on 
the physical process being studied, e.g., $e^{i{\bf q} \cdot {\bf r}}$ for non-resonant inelastic 
x-ray scattering (NRIXS) or the expansion $(\hat{\bf e} \cdot {\bf r}) + (i/2)(\hat{\bf e} \cdot {\bf r})({\bf q} \cdot {\bf r}) + \dots$ 
for x-ray absorption (XAS); $\hat{\bf e}$ being the photon polarization vector.  
Using the Bethe-Salpeter Hamiltonian, the Green's function for the excitation can be 
approximated in a two-particle form as
~
\begin{equation}
G(\omega) = [\omega - H_{\rm BSE}]^{-1}
\end{equation}

\noindent where the Bethe-Salpeter Hamiltonian is typically given by
~
\begin{equation}
H_{\rm BSE} = H_{\rm e} - H_{\rm h} - V_{\rm D} + V_{\rm X} .
\label{H_BSE}
\end{equation}

\noindent The term for the core-hole
~
\begin{equation}
H_{\rm h} = \epsilon_{\rm c} + \chi - i \Gamma
\end{equation}

\noindent contains the average core-level energy $\epsilon_{\rm c}$, the spin-orbit interaction $\chi$, 
and the core-hole life-time $\Gamma$.  In most practical work, the excited electron Hamiltonian
~
\begin{equation}
H_{\rm e} = H_{\rm KS} + \Sigma^{\it GW} - V_{xc}
\end{equation}

\noindent is approximated by the Kohn-Sham Hamiltonian $H_{\rm KS}$ with a {\it GW} self-energy 
correction $\Sigma^{\it GW}$ where the exchange-correlation energy, $V_{xc}$, is subtracted off 
due to double-counting.  The excited electron and hole interact via the Coulomb interaction 
within the mean-field of the remaining electrons.  This is separated into the attractive direct term
%~
%\begin{equation}
%V_{\rm D} = \hat{a}_{\rm v}^{+}({\bf r}, \sigma) \hat{a}_{\rm h}({\bf r^{\prime}},
%\sigma^{\prime}) 
%W({\bf r},{\bf r^{\prime}},\omega) 
%\hat{a}_{\rm v}({\bf r}, \sigma) \hat{a}_{\rm h}^{+}({\bf r^{\prime}}, \sigma^{\prime}) ,
%\end{equation}
~
\begin{equation}
V_{\rm D} = \hat{a}_{\rm v}^{+}({\bf r}, \sigma) \hat{a}_{\rm c}({\bf r^{\prime}},
\sigma^{\prime}) W({\bf r},{\bf r^{\prime}},\omega)
\hat{a}_{\rm v}({\bf r}, \sigma) \hat{a}_{\rm c}^{+}({\bf r^{\prime}}, \sigma^{\prime}) ,
\end{equation}

\noindent which is screened by the other electrons in the system, and the repulsive exchange term
%~
%\begin{equation}
%V_{\rm X} = \hat{a}_{\rm v}^{+}({\bf r},\sigma) \hat{a}_{\rm h}({\bf r^{\prime}},\sigma^{\prime}) 
%\frac{1}{|{\bf r}-{\bf r^{\prime}}|} 
%\hat{a}_{\rm v}({\bf r^{\prime}},\sigma) \hat{a}_{\rm h}^{+}({\bf r},\sigma^{\prime}) ,
%\end{equation}
~
\begin{equation}
V_{\rm X} = \hat{a}_{\rm v}^{+}({\bf r},\sigma) \hat{a}_{\rm c}({\bf r},\sigma)
\frac{1}{|{\bf r}-{\bf r^{\prime}}|}
\hat{a}_{\rm v}({\bf r^{\prime}},\sigma^{\prime}) \hat{a}_{\rm c}^{+}({\bf r^{\prime}},\sigma^{\prime}) ,
\end{equation}

\noindent which is treated as a bare interaction.  The $\hat{a}_{\rm v}^{+}$ ($\hat{a}_{\rm v}$) 
operator creates (annihilates) an electron in the valence level while 
$\hat{a}_{\rm c}^{+}$ ($\hat{a}_{\rm c}$) creates (annihilates) an electron in a core level.  

Our implementation of the {\it GW}-BSE method is referred to as {\sc ocean} 
(Obtaining Core Excitations from {\em Ab initio} electronic structure and NBSE)
and we have previously presented it in detail \cite{vinson.ocean}; NBSE refers to the NIST 
BSE solver.  Because solving the Bethe-Salpeter equation at the level of approximation described 
above is computationally intensive compared to other methods of calculating x-ray spectra our approach 
makes several reasonable approximations to improve the efficiency of the calculation.  Specifically, 
within a plane-wave approach to solving the Kohn-Sham equations, we use pseudopotentials to reduce 
the number of electrons and size of the plane-wave basis.  The {\it GW} self-energy 
is obtained through the highly efficient many-pole self-energy approximation when appropriate 
\cite{kas.mpse}.  The effort required to obtain the screening for the direct 
interaction is also reduced by utilizing a hybrid real-space approach in which the screening 
response is evaluated at the RPA level locally about the absorbing site, but the long range 
screening is approximated with a model dielectric function \cite{shirley.screening}.

Despite these efficacious strategies, the largest system treated with our previous 
implementation of {\sc ocean} was a water cell consisting of 17 molecules \cite{vinson.h2o}.  
To extend the capabilities 
of {\sc ocean} to treat larger systems we have made several improvements.  
A calculation with {\sc ocean} consists of four stages
\begin{enumerate} \itemsep1pt \parskip0pt \parsep0pt
\item DFT \quad
2. Translator \quad
3. Screening \quad
4. BSE
\end{enumerate} 
where stage 2 is a translation layer that allows different DFT packages to be used as the foundation for {\sc ocean}. 

The limiting points of the 
calculation previously were stages 1 and 3, solving the Kohn-Sham equations and evaluating the screening response 
to the core-hole.  Stage 4, the actual evaluation of the Bethe-Salpeter Hamiltonian, was also a limiting 
point for systems that required sampling numerous atomic sites.  Therefore, our efforts focused 
on (i) improving the efficiency of the DFT calculation and (ii) parallelizing the evaluation of 
the screening response and the Bethe-Salpeter Hamiltonian.

\section{Implementation}

\subsection{Optimal Basis Functions}

Our previous version of {\sc ocean} used {\sc abinit} \cite{abinit0, abinit1, abinit2, abinit3} as the DFT solver.
{\sc ocean} may now be alternatively based on wavefunctions obtained with Quantum{\sc espresso} \cite{espresso1}.
For the purpose of this paper, we report results based on the use of Quantum{\sc espresso} rather than
{\sc abinit}, though either DFT solver may be chosen depending on the preference of the user.

The loss function in eqn.~\ref{loss.fcn} is calculated by transforming the implied integral over all space
into sums over reciprocal-space $\mathbf{k}$-points within the Brillouin zone and 
real-space $\mathbf{x}$-points within the unit cell, as is standard in calculations of periodic systems. 
The sum over $\mathbf{k}$-points requires a denser mesh for converging spectra than properties such 
as the density. Additionally, the BSE approach necessitates summing over a large number of unoccupied
states which are also not needed when looking at ground-state properties.  Thus, a considerable 
number of Kohn-Sham states must be constructed.

We reduce the computational cost of generating the Bloch functions through the use of 
optimal basis functions (OBF) \cite{shirley.obf}.  We have implemented the OBF routines of Prendergast
and Louie as a middle-layer in the {\sc ocean} code \cite{prendergast.obf}. The OBFs are a method of 
$\mathbf{k}$-space interpolation and basis reduction. A fully self-consistent DFT calculation is carried out to converge the
density, using only enough bands to cover the occupied states.  With
this density a non-SCF calculation is performed, including
unoccupied bands for the screening and BSE calculations
(the density is held constant and the Kohn-Sham eigensystem is solved for all the bands needed for the BSE). This
second calculation is used as a basis to create the OBFs.
By using the OBFs we achieve a significant reduction in the time spent calculating 
the Bloch functions for a given system.  Further details and quantitative results are 
presented in section \ref{results.obf}.

\subsection{BSE Hamiltonian}
\label{bseHam}

In {\sc ocean} the BSE Hamiltonian acts on a space containing a core-level hole with index 
$\alpha$ and a conduction band electron with indicies $n,\mathbf{k}$. A vector in this space 
is described by the coefficients $\psi_{\alpha,n,\mathbf{k}}$, and the photoelectron wavefunction 
for a given core index $\alpha$ is easily expanded in real space from the conduction-band Bloch functions
\begin{align}
\vert \Phi_\alpha (\mathbf{x},\mathbf{R}) \rangle = \sum_{n,\mathbf{k}} \psi_{\alpha,n,\mathbf{k}} \, e^{i \mathbf{k} \cdot (\mathbf{R+x})} \vert u_{n,\mathbf{k}}(\mathbf{x}) \rangle .
\end{align}
%The action of the BSE Hamiltonian is broken up into short-range and long-range components, and 
%different parallelization strategies are used for each.
%
Within OCEAN we are interested primarily in the resulting x-ray spectrum, rather than obtaining the 
actual eigenvalues and eigenvectors of the Bethe-Salpeter Hamiltonian, and so the iterative Haydock 
technique is used \cite{Haydock,PhysRevB.59.5441}. 
The advantage of an iterative technique is that the complete Hamiltonian does not need to be 
explicitly constructed or stored. Rather, at each step in the Haydock scheme the BSE Hamiltonian 
acts on the vector determined by the previous iteration,
\begin{align}
\label{haydock1}
\psi^{i+1}_{\alpha,n,\mathbf{k}} = H_\text{BSE}\, \psi^{i}_{\alpha,n,\mathbf{k}},
\end{align}
where $i$ gives the $i$th vector, 
and convergence of the spectrum can be achieved in only a few hundred iterations 
(true diagonalization requires a number of steps equal to the dimension of the matrix).
A more explicit explanation is given by Benedict and Shirley in Ref.\  \citenum{PhysRevB.59.5441}. 
We divide up equation \ref{haydock1} by separating the BSE Hamiltonian into pieces, 
each of which is evaluated in its ideal or most compact basis. 
This is outlined in the following two sections. 
The full vector $\psi^{i+1}$ is then the sum of contributions from each piece.

\subsubsection{Long-range}

Only the direct interaction has a long-range component as the exchange term requires both $\mathbf{r}$ 
and $\mathbf{r'}$ be at the core-hole site. The screened-coulomb interaction can be expanded 
in spherical harmonics and separated by angular momentum $l$,
%with each angular momentum $l$ picking up a factor of $r_>^{-(l+1)}$.  
\begin{align}
W(\mathbf{r},\mathbf{r}') &= \int \!\! d\mathbf{r}'' \frac{ \epsilon^{-1}(\mathbf{r},\mathbf{r}'') }{\vert \mathbf{r}'' - \mathbf{r}' \vert}  = \sum_{l=0}^{\infty} W_l(\mathbf{r},\mathbf{r}') .
\end{align}
The long-range part includes only the $l=0$ term, 
\begin{align}
W_0(\mathbf{r},\mathbf{r}') &= \int \!\! d\mathbf{r}''  \frac{ \epsilon^{-1}(\mathbf{r},\mathbf{r}'') }{\left[r_>\right]_{\mathbf{r}'',\mathbf{r}'}},
\end{align}
that goes as $1/r$ away from the core hole. All higher $W_l$ are treated as short ranged. 
By integrating out the core-hole dependence via the core-hole density $\rho_\alpha$  
\begin{align}
W_0(r) &= \int \!\! d\mathbf{r}' \rho_\alpha(\mathbf{r}') \int\!\! d\mathbf{r}''  \frac{ \epsilon^{-1}(\mathbf{r},\mathbf{r}'') }{\left[r_>\right]_{\mathbf{r}'',\mathbf{r}'}},
\end{align}
the long-range component of the BSE Hamiltonian is a function of the electron spatial coordinate only.

We evaluate the action of $W_0$ on a vector by first transforming to a super-cell space
\begin{align}
\phi_\alpha(\mathbf{x},\mathbf{k}) &= \sum_n \psi_{\alpha,n,\mathbf{k}} \, e^{i \mathbf{k} \cdot \mathbf{x}} \vert u_{n,\mathbf{k}}(\mathbf{x}) \rangle  \\ \nonumber
\phi_\alpha(\mathbf{x},\mathbf{R}) &\equiv \mathcal{F}_{\mathbf{k} \rightarrow \mathbf{R}} [ \phi_\alpha(\mathbf{x},\mathbf{k}) ] ,
\end{align}
where $R$ are the lattice vectors and $\mathcal{F}$ indicates a Fourier transform. The core-hole 
index $\alpha$ can refer to different angular momentum and spin $l m \sigma$ states of the core hole, but the long-range component of the direct term does not mix spin or angular momentum, and so these are treated sequentially. The $\mathbf{k}$-space grid used in an {\sc ocean} calculation defines a maximum range of the screened core-hole potential. 

\begin{algorithm}
\caption{Computing $\psi^{i+1} = W_0 \psi^{i} $}
\label{W0}
\begin{algorithmic}[1]

	%\State $\psi_{n,\mathbf{k}}^{i+1} = 0$
    \For {each $x$ and $\alpha$}
	\State $\phi(\mathbf{k}) = \sum_n \psi_{n,\mathbf{k},\alpha}^{i} \,e^{i \mathbf{k} \cdot \mathbf{x}} u_{n,\mathbf{k}}(\mathbf{x})$
	\State $\phi(\mathbf{R}) = \mathcal{F}_{\mathbf{k} \rightarrow \mathbf{R}} [ \phi(\mathbf{k}) ]$
	\State $\phi(\mathbf{R}) = W(\mathbf{x},\mathbf{R}) \times \phi(\mathbf{R})$
	\State $\phi(\mathbf{k}) =   \mathcal{F}_{\mathbf{R} \rightarrow \mathbf{k}} [ \phi(\mathbf{R}) ]$
	\State $\psi_{n,\mathbf{k},\alpha}^{i+1} \pluseq \phi(\mathbf{k}) \,e^{-i \mathbf{k} \cdot \mathbf{x}} u^{*}_{n,\mathbf{k}}(\mathbf{x})$
	\EndFor

\end{algorithmic}
\end{algorithm}

The operation $\psi^{i+1} = W_0 \psi^{i}$ is laid out in algorithm \ref{W0}. Since the direct 
interaction is diagonal in real-space this procedure is easily parallelized by distributing the 
$\mathbf{x}$-points among all the processors. The parallel algorithm has the additional final step 
of summing the vector $\psi^{i+1}$ over all the processors. This limits the scaling, but the vector 
itself is not large. 
%The scaling is limited by the parallel summation of $\psi^{i+1}$, but the vector is itself not large.  
Both the number of $\mathbf{x}$-points 
and the number of empty bands required for a given energy range scale as the volume of the system.

\subsubsection{Short-range}

The short-range components of the Hamiltonian are calculated by projecting the conduction-band states 
into a local basis around the core hole. These basis functions have a well-defined angular momentum 
around the absorbing atom and are reasonably complete such that near the atom
\begin{align}
\label{local_basis}
\varphi_{n,\mathbf{k}}(\mathbf{r})\vert_{r<r_c} = \sum_{\nu,l,m} A_{n,\mathbf{k}}^{\nu,l,m} R_{\nu,l}(r) Y_{l,m}(\hat{r}) .
\end{align}
$Y_{l,m}$ are the usual spherical harmonics and, following the
ideas of the projector augmented wave method, $R_{\nu,l}$ are taken to be
solutions to the isolated atom \cite{PhysRevB.50.17953,Shirley200477}. 
This both allows us to capture the correct, 
oscillatory behavior of the valence and conduction states near the core-hole and gives us a compact 
basis for calculating the matrix elements of the short-range direct and exchange interactions. In 
practice, our core-hole will typically have 1 ({\it s}) or 3 ({\it p}) angular momentum states, 
and our conduction electrons will have 4 projectors each for $l=s, \cdots, f$ or 64 {\it lm} states. 
For an L edge this gives a maximum matrix dimension of $4\times3\times64=768$, including spin degrees 
of freedom for both the electron and the hole. This is completely independent of the overall system size.

The time-consuming steps in computing the short-range interactions are projecting into and out of 
the localized basis which requires summing over all of the bands and $\mathbf{k}$-points. The 
mapping of band states to localized states is precomputed and stored. We expand the exchange 
and local direct by angular momentum and exploit selection rules to limit the number of multipole 
terms. The various pieces of the short-range Hamiltonian are distributed to different processors. 
The small number of multipole terms and size disparity between them limits the degree of parallelization, 
but alleviates the need for a communication step. 
Each processor adds its results for the long- and short-range interactions, and then a single 
synchronizing summation of $\psi^{i+1}$ is carried out. 

\subsection{Screening}
\label{screen}

In the direct interaction the core-hole potential is screened by the dielectric response of the system.  
We calculate this response within the random phase approximation (RPA)
\begin{align}
\chi^0(\mathbf{r},\mathbf{r}',\omega) = \int \!\! \frac{\,d\omega'}{2 \pi i} G(\mathbf{r},\mathbf{r}',\omega')G(\mathbf{r}',\mathbf{r},\omega'-\omega).
\label{chi_rpa}
\end{align}
We evaluate this expression in real-space around the core-hole, and, taking advantage of the localized nature of 
near-edge core excitations, we limit our full calculation to a sphere with a radius of approximately 
$r=8$~a.u., splicing on a model dielectric function for the long-range behavior \cite{shirley.screening}. 
The cross-over radius from RPA to model is a convergence parameter so that for each material one may 
ensure that this approximation has no discernible effect on the calculated spectra. We use static screening $\omega=0$ which assumes 
that the exciton binding energy is small compared to the energy scale for changes in the dielectric 
response, {\it i.e.}, smaller than the band gap. 

Our real-space grid is 900 points determined from a 36 point angular grid and 25 uniformly spaced radial 
points.  We carry out the integral over energy in eqn.\ \ref{chi_rpa} explicitly along the imaginary axis, 
and for large systems the bulk of time is spent projecting the wavefunctions onto this grid and constructing the Green's function 
\begin{align}  \label{greens.fcn}
G(\mathbf{r},\mathbf{r}',\mu + it) = \sum_{n,\mathbf{k}} \frac{\psi_{n,\mathbf{k}}(\mathbf{r}) \psi^{*}_{n,\mathbf{k}}(\mathbf{r}')}{\mu + it - \epsilon_{n,\mathbf{k}}},
\end{align}
where $\mu$ is the chemical potential. 
Using the OBFs, we determine the Bloch functions on the spherical grid and distribute the calculation 
across processors by dividing up the spatial coordinates.  On modern computer architectures this operation 
will be limited by memory bandwidth. To alleviate this we have each processor work on one or more small 
blocks of the Green's function such that we can be confident that the block will remain in 
the cache during the summation over bands. 

To gain an additional level of parallelization we can also distribute the calculation by $\mathbf{k}$-point.  
We divide the total processors into pools of equal size, and 
every pool calculates $G_\mathbf{k}$. The sum $G = \sum_\mathbf{k} G_\mathbf{k}$ is 
carried out across the pools, placing the complete $G$ on the master pool. 
 Then each processor in the master pool calculates $\chi^0$ via equation \ref{chi_rpa}, 
 performing the integral over the imaginary energy axis for its blocks.  
 Finally, the complete matrix $\chi^0$ is written to disk.  In 
practice, a $2\times2\times2$ shifted $\mathbf{k}$-point grid is used for screening calculations, giving 
8 unique points in the Brillouin zone and allowing up to 8 pools. There is some cost to performing the 
parallel sum of $G$ over pools and inefficiency in evaluating $\chi^0$ on only the master pool, however neither of 
these contribute significantly to the running time.

~
\begin{table}[h]
\centering
%\caption{Timing budget variation with system size.  The central portion 
%of the table lists the percentage of total run-time spent on each stage 
%of the calculation.  The final line shows the total run-time for each 
%supercell relative to the run-time for the conventional cell.  The 
%number of processors listed refers to the DFT stage of the calculation only.
%The remaining stages were carried out on a single core.}
%\begin{tabular*}{\columnwidth}{c @{\extracolsep{\fill}} cccccc}
%  \hline
%  \hline
%  & Supercell (N) & 1 & 2 & 3 & 4 \\
%  \hline
%  & Rel.~System Size & 1 & 8 & 27 & 64 \\
%  \hline
% & \# Processors & 8 & 32 & 64& 128 \\
%  \hline
%  \hline
%  & DFT & 38.0 & 81.8 & 65.7 & 43.6 \\
%  & Translator & 7.3 & 10.9 & 10.7 & 15.1 \\
%  & Screening & 29.4 & 4.1 & 1.8 & 2.2 \\
%  & BSE & 25.3 & 3.2 & 21.8 & 39.1 \\
%  \hline
%  \hline
%  & Rel.~Total Time  & 1 & 10.1 & 123.3 & 559.9 \\
%  \hline
%  \hline
%\end{tabular*}
%\end{table}
\caption{Timing budget variation with system size.  The central portion
of the table lists the total run-time in minutes spent on each stage
of the calculation.  The DFT portion of the calculation was performed 
in parallel and the time shown is the run-time multiplied by the number 
of cores used.  The remaining stages were carried out on a single core.}
\begin{tabular*}{\columnwidth}{c @{\extracolsep{\fill}} cccccc}
  \hline
  \hline
  & Supercell (N) & 1 & 2 & 3 & 4 \\
  \hline
  & Rel.~System Size & 1 & 8 & 27 & 64 \\
  \hline
  \hline
  & DFT & 22 & 832 & 17161 & 265226 \\
  & Translator & 0.08 & 2.5 & 43.8 & 754 \\
  & Screening & 0.2 & 2.0 & 15.4 & 76.5 \\
  & BSE & 0.5 & 26.1 & 250 & 1915 \\
  \hline
  \hline
\end{tabular*}
\end{table}

\section{Results}

To demonstrate the scaling performance of {\sc ocean} we consider the Ti
L-edge XAS of supercells of SrTiO$_{3}$.  Pure SrTiO$_{3}$ is a wide band-gap
semiconductor and incipient ferroelectric that assumes an undistorted
cubic structure at ambient conditions.  SrTiO$_{3}$ is a foundational
material on which a remarkable variety of electronic and
opto-electronic devices are based.  Epitaxially interfacing SrTiO$_{3}$
with other oxide compounds, notably LaAlO$_{3}$, can yield interfacial
2-dimensional electron gases that are good conductors \cite{hwang.alo-sto}, 
superconducting \cite{reyren.alo-sto}, show magnetic ordering \cite{brinkman.alo-sto}, 
or that are even simultaneously superconducting and magnetically ordered 
\cite{ashoori.alo-sto, bert.alo-sto}.  Doping SrTiO$_{3}$ also produces a wide range 
of useful physical properties.  Addressing these interesting systems numerically 
requires the use of large supercells.  For instance, a doping level
of 3.7 $\%$ (1/27) necessitates a $3\times3\times3$ supercell while for a 1.56 $\%$
(1/64) doping level a $4\times4\times4$ supercell must be constructed.

In this section we consider run-times for cubic supercells of pure SrTiO$_{3}$ 
with N = $\{1,2,3,4\}$ repetitions of the conventional cell in each direction.
This corresponds to systems with $\{5,40,135,320\}$ atoms and $\{32,256,864,2048\}$ 
valence electrons, respectively.  
We begin in section \ref{results.serial} with benchmark calculations by investigating the 
single-core time-scaling of the original screening and BSE portions 
of the calculation.  The four supercells are run with equivalent 
parameters and the usual basis of Kohn-Sham orbitals, that is, without 
employing the optimal basis functions.  We next consider in section \ref{results.obf} the 
savings gained through the {\bf k}-space interpolation scheme of optimal 
basis functions.  In sections \ref{results.bse} and \ref{results.screening}, 
we determine the scaling behavior 
of the parallel implementation of the BSE and screening routines, respectively.  
We end in section \ref{results.examples} with a brief demonstration of {\sc ocean} by calculating 
the x-ray absorption and emission spectra of an organic molecule commonly used in light 
emitting diode devices.

\subsection{Single Processor Calculations}  \label{results.serial}

Before discussing the improvements we have made it is informative to 
establish the prior baseline performance of {\sc ocean}.  We performed a 
series of timing tests for which the DFT portion of the calculation 
was executed in parallel over a given number of cores while the remaining 
stages of the calculation used only a single core.
We use the Quantum{\sc espresso} density-functional theory package to
generate the ground-state electronic structure upon which the spectral
calculations are based.  These calculations are performed with
norm-conserving pseudopotentials obtained from the {\sc abinit} 
distribution \cite{abinit0, abinit1, abinit2, abinit3}
with the exception of Ti for which we made a pseudopotential with
semi-core states included in the valence configuration.  We employ the
local-density approximation to the exchange correlation functional and
truncate the planewave basis at 50 Ry.  $\Gamma$-point sampling was used
to obtain the ground-state electron density, which was subsequently
expanded into separate sets of Kohn-Sham orbitals for the evaluation of the
screening and the Bethe-Salpeter Hamiltonian.  In each case, the
number of empty bands utilized equaled the number of occupied bands.
For the screening response, states were constructed at a single k-point while to 
evaluate the Bethe-Salpeter Hamiltonian states were expanded on a shifted 
$2\times2\times2$ grid.  The above values were sufficient for convergence of 
the final spectrum for supercells with N $\ge$ 2.  Since our purpose at 
present is to study the scaling performance of {\sc ocean} we use these 
same values for the N = 1 case (conventional cell) even though this {\bf k}-point 
sampling is insufficient for convergence at this size.  (Convergence at N = 1 can be
reached by increasing the {\bf k}-point sampling to $2\times2\times2$ for the density
and evaluation of the screening, and to $3\times3\times3$ for the BSE.)

%Table 1 reports the timing budget separated by stage of the calculation for 
%each of the four supercells.  For very small systems roughly equal time is 
%spent on the DFT, screening and BSE portions of the calculation.  As the 
%system size increases, the time required for the DFT and BSE steps 
%dominate that required for the screening calculation.  However, only a single 
%Ti site is sampled for each supercell.  In cases for which multiple atomic 
%sites must be evaluated the times for the screening and BSE stages will increase 
%proportionally.  Thus, while effort should clearly be extended to increase the 
%efficiency of the DFT and BSE portions of the calculation, parallelizing 
%the calculation of the screening response is also worthwhile.  Table 1 further
%provides the total run-time for each supercell normalized to the run-time 
%for the conventional cell.  Of course, the time required for the DFT calculations 
%depends on the number of cores used, but these specific results convey a 
%practical impression of the scaling for typical numbers of cores.

Table 1 reports the timing budget separated by stage of the calculation for each of 
the four supercells.  The DFT segments of the calculations were run in parallel and 
the time reported is the product of the run-time and the number of cores used.  The 
remaining stages of the calculation were all run on a single core; only a single Ti 
site was interrogated for each supercell.  Extrapolating these times to all Ti sites 
in each supercell, we find that the run-time required for the three stages after the 
DFT portion constitutes approximately 30 $\%$ of the total calculation time for the 
larger supercells.  Parallelization of these segments could then reduce their run-time 
to a small fraction of the DFT run-time.  Table 1 shows that the time spent in the BSE 
stage dominates that used in the screening portion of the calculation.  Thus, it is 
essential to implement an effective parallelization scheme for the BSE stage; this will 
be demonstrated below.  The wavefunction translation step also becomes time consuming 
for large systems.  At present, we have not sought to improve the efficiency of this 
process, but future efforts may be directed at reducing the time required here.  

~
\begin{figure}[h*]
  \includegraphics[angle=270,width=9.0cm]{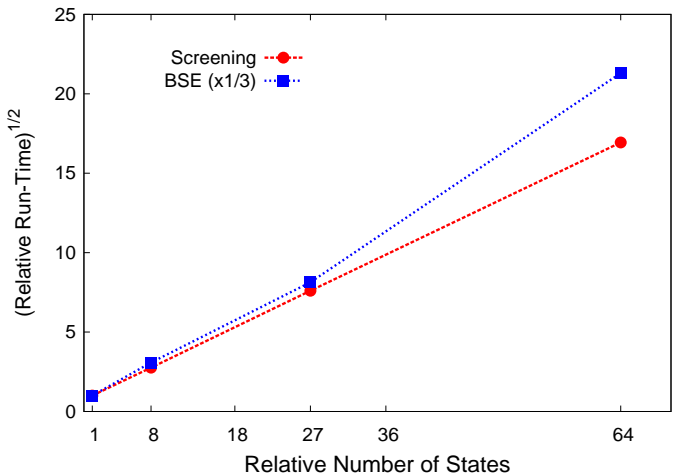}
  \caption{{\bf Run-Time Scaling.} The relative run-times for the
screening and BSE portions of the calculation are presented versus
system size.  The screening (left axis, red circles) scales with the system
size to the second power and the BSE (right axis, blue squares) goes as the
system size to the fourth power.}
 \label{scaling}
\end{figure}

The individual run-time scaling of the screening and BSE
stages of the calculation are presented in Figure 1.  These results  
demonstrate that both the screening
calculation and the evaluation of the BSE 
scale as the number of states to the second power. 
Since we use the same k-point grids for all supercells the
variation in the number of states comes from the increasing number of
bands 
%, plane-waves in the DFT, and $\mathbf{x}$-points in the BSE; 
%all 
which grows proportionally with the supercell size. 
In this example we have considered a single site in the super cell. 
The number of sites also grow with volume, 
leading to an overall scaling of volume cubed for both the screening and BSE. 
Given that as the system size increases the BSE stage comes to dominate the 
overall run-time, it is clearly advantageous to find an effective parallelization 
scheme for this stage.  This is discussed below in section \ref{results.bse}.

From eqn.\ \ref{greens.fcn} it appears that the screening calculation should scale directly 
with the number of bands as the radial grid is independent of system size. 
However, for large systems much 
of the time is spent projecting the DFT wave functions onto the radial grid 
from the planewave basis. Both the number of planewaves and bands grow 
with system size leading to this second power behavior. 
This indicates that an improved approach to projecting the Kohn-Sham states onto 
the local basis is an avenue to further reducing the time spent in the 
screening routine.

Per section 3.2, the BSE Hamiltonian is broken into three sections: 
non-interacting, short-range, and long-range. The non-interacting term is diagonal, and, 
for a single site scales directly with system size. The short-range part, like the screening 
calculation, relies on a small, localized basis that does not change with system size. 
Also like the screening calculation, for large system the projection of the DFT orbitals into this 
localizated basis to determine the coefficients $A$ (eqn.\ \ref{local_basis}) will grow with both 
the number of planewaves and bands. While this projection happens only once at the beginning 
of the BSE stage, it does lead to scaling with the second power of the system size. 
The long-range part of the BSE is the most computationally expensive. 
As shown in algorithm 1, the action 
of $W_0$ grows with both the number of  {\bf x}-points and the number of empty 
states, both of which scale linearly with system size. Therefore the BSE section of the 
code is expected to scale with the second power of system size, as is confirmed in 
figure \ref{scaling}.
%The scaling of the BSE as the system size to the fourth power is 
%as expected.  The size of the vector in the Bethe-Salpeter 
%Hamiltonian is the product of the number of {\bf x}-points and the number of empty 
%states, both of which scale linearly with system size.  Acting on this 
%vector requires operations that scale as the size of the vector squared, 
%or to the fourth power of the system size in this case.

\subsection{Reduced Basis}  \label{results.obf}

For the SrTiO$_{3}$ supercells presented in this work, $\Gamma$ point sampling was used as an 
input to the OBF scheme and the Bloch functions were interpolated onto $2\times2\times2$ $\mathbf{k}$-point 
grids (For the single cell a $2^3$ $\mathbf{k}$-point grid was interpolated to a $3^3$ grid). 
As NSCF DFT calculations scale linearly with the number of $\mathbf{k}$-points this represents 
a potential 8x speedup (3.4x for the conventional cell).  This gain, however, is partially offset 
by the additional steps needed to carry out the OBF interpolation. 

\begin{table}[h]
\centering
\caption{The relative time required for each step in generating Bloch functions and the speedup with respect to the total run time achieved by using the OBFs (The total run-time includes steps not explicitly listed in the table).}
\begin{tabular*}{\columnwidth}{c @{\extracolsep{\fill}} cccccc}
  \hline
  \hline
  & Supercell (N) & 1 & 2 & 3 & 4 \\
  \hline
  & Rel.~System Size & 1 & 8 & 27 & 64 \\
  \hline
 & \# Processors & 8 & 32 & 64& 128 \\
  \hline
  \hline
  & SCF & 0.48 & 2.20 & 6.45 & 42.5 \\
  & NSCF & 0.06 & 0.38 & 7.11 & 38.6 \\
  & OBF & 0.18 & 1.55 & 12.1 & 49.7 \\
  \hline
  & Total  & 1.00 & 4.90  & 30.3 & 152 \\
\hline
& Speedup & 1.27x & 1.22x & 2.24x & 2.46x \\
  \hline
  \hline
\end{tabular*} 	
\end{table}

In Table 2 we show the relative time for the SCF, NSCF, and OBF stages that are needed to 
generate Bloch functions for the BSE calculation. By assuming that the NSCF would necessarily 
take 8x (3.4x) longer without the OBF interpolation we estimate the savings as a percentage of 
the total run-time. While the small cells show only a modest improvement, the $3^3$ and $4^3$ 
cells complete in less than half the time when using the OBF scheme. Generically, the expected savings will 
depend most strongly on the needed $\mathbf{k}$-point sampling for the system under investigation. 
Metallic systems require much denser $\mathbf{k}$-point grids for convergence and will yield 
correspondingly larger savings.

\subsection{BSE Scaling}  \label{results.bse}

To investigate the scalability of our parallel BSE solver with processor number we focus on the $4\times4\times4$ 
supercell of SrTiO$_{3}$.  This system approaches the limits of single node execution due to memory 
considerations.  A significant amount of time for each run is spent reading in the wavefunctions.  For 
this example the wavefunctions require 12~GB of space (3072 conduction bands, 8 ${\mathbf k}$-points, and 32,768 
${\mathbf x}$-points).  The time needed to read these data typically ranged from 40-60~seconds \cite{diskio}.  Currently, 
the wavefunctions are read in by a single MPI task and then distributed so this time is relatively constant 
with processor count.  To give a better picture of the scaling, we subtract out the time required for 
this step before comparing runs. If multiple runs are carried out on the same cell, varying atomic site, 
edge, or x-ray photon (polarization or momentum transfer), the wavefunctions are kept in memory, and 
therefore this upfront cost will be amortized over all the calculations. In the present example we 
calculate the spectra for three polarizations, using 200 Haydock iterations. 

\begin{figure}[h*]
  \includegraphics[angle=270,width=9.0cm]{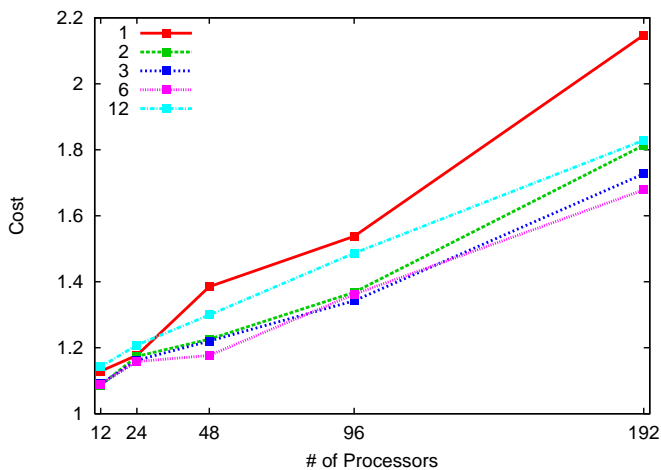}
  \caption{{\bf BSE Cost.} The cost, ratio of actual run-time to theoretical linear scaling, as a function of processor count. Each line represents a different number of threads per node. One or two MPI tasks per socket (6 or 3 threads) are seen to give the best performance. }
 \label{BSE_scale}
\end{figure}

To assess the scaling of the BSE section we use the metric cost. The cost reflects what the user would be charged on a computing facility: the run-time multiplied by the number of processors. 
%[The motivation for this representation is twofold; 1) It provides better granularity when plotted compared to speedup or run-time which would both change by 2 orders of magnitude as we changed from 1 to 192 processors, and 2) It better communicates to end-users the trade-off between running for a longer on fewer processors versus for a shorter time on more] (My preference is to omit this sentence; it can be put in the referee response, though.). 
A perfectly parallelized code would maintain a cost of 1.0 when run across any number of processors. 
Costs less than 1.0 would indicate some superscaling behavior, most likely from accidental cache reuse. 
% [and, with proper refactoring, these gains could be realized for runs using fewer processors as well] (I don’t know how many people will understand this). 
In figure \ref{BSE_scale} we show the relative cost of the BSE section as a function of the number of processors. The testbed for this section consists of identical, dual socket nodes with 6 processors (cores) per chip and connected via high-speed interconnects. Each multi-node test was run five times and the best time for each was used.

The BSE section is a hybrid MPI/OpenMP code so we compare execution with various levels of threading. 
As can be seen in figure \ref{BSE_scale}, the use of threads improves the speed of the calculation over 
pure MPI. This is true even when 12 threads are used, requiring communication across the two sockets via 
OpenMP. We also see very acceptable scaling with processor count. The benchmark calculation takes a little 
over 8.5~hours on a single node and single thread, while running on 192 cores can be done in approximately 
3.5~minutes for a real-world speedup of 114x. 

Sublinear scaling, a cost in excess of $1.0$, is, in general, the result of non-parallelized sections of code, synchronization costs, and memory bandwidth constraints. With the exception of the aforementioned wavefunction initialization, the BSE code has very little explicit serial execution. Further work is needed to investigate and alleviate the bottlenecks preventing linear scaling.

\subsection{Core-hole Screening Scaling}  \label{results.screening}

In this section we investigate the timing of calculating the valence electron screening of the core-hole. 
As stated before, for core-level spectroscopy we are mainly interested in the local electronic response. 
Limiting our calculation of the RPA susceptibility to a region of space around the absorbing atom makes the 
calculation much cheaper than traditional planewave approaches without sacrificing accuracy 
\cite{shirley.screening}.  Within this approach, and for most other methods of calculating 
core-excitation spectra, a separate screening calculation is required for each atomic site.  
In systems with inequivalent sites due to differences in bonding, defects, or vibrational 
disorder contributions from each atom must be summed to generate a complete spectrum. 

\begin{figure}
  \includegraphics[angle=270,width=9.0cm]{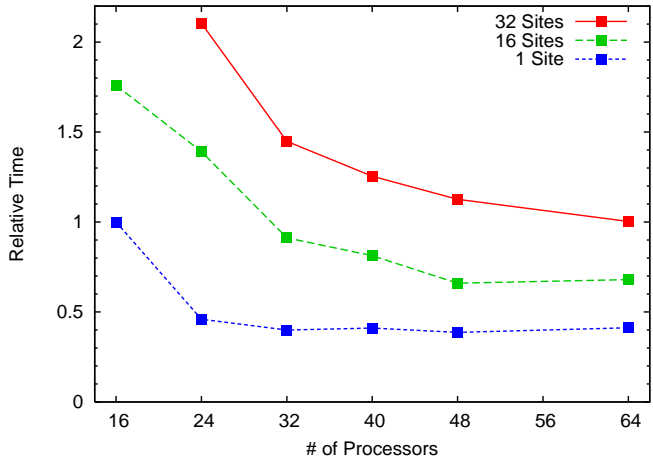}
  \caption{{\bf Screening scaling.} The relative time for the screening section as a function of the number of processors. We show both the total time required for a single site, 16 sites, and 32 sites relative to the time needed to run the single site on 16 processors. 
The per site time is lower with a larger number of sites due to amortization of i/o and other initialization costs. While the scaling with processor number is poor, we see only very limited detrimental effects from using more processors, and therefore the screening calculation is run using the same processor count as the other stages.  }
 \label{Screen_scale}
\end{figure}

To test the behavior of the screening section we use the same test case as for the BSE, the 
$4\times4\times4$ supercell 
of SrTiO$_{3}$. The RPA susceptibility is calculated using a single $\mathbf{k}$-point and 4096 bands. We show 
results for a single site, 16 sites, and 32 sites (out of the 64 titanium atoms in our supercell). The 
number of sites will vary based on the system being investigated. Impurities may 
require only a single site, but liquids or other disordered systems necessitate averaging over 
many sites \cite{vinson.h2o,PhysRevB.90.205207}. 
 As is evident in figure \ref{Screen_scale}, this section of the calculation does not scale beyond a few 
dozen processors. This is especially true when only investigating a single site. While better scaling is 
desirable, as observed in Table 1 the total time for this section is quite small.   In this particular 
test the screening calculation took just under 8~min for a single site while the initial DFT calculations
required approximately 3~hours on 128~processors. 

\subsection{Example Spectra}  \label{results.examples}

%Figure \ref{Ti_L-edge} presents the Ti L edge of pure SrTiO$_{3}$ calculated for both the 
%conventional cell and the $5\times5\times4$ supercell.  We have previously reported the Ti 
%L edge of SrTiO$_{3}$ obtained from {\sc ocean} \cite{vinson.ocean} and figure \ref{Ti_L-edge} 
%simply serves to verify that the fidelity of the computational scheme has been preserved 
%through the modifications we have made.  

\begin{figure}
\includegraphics[angle=270,width=9.0cm]{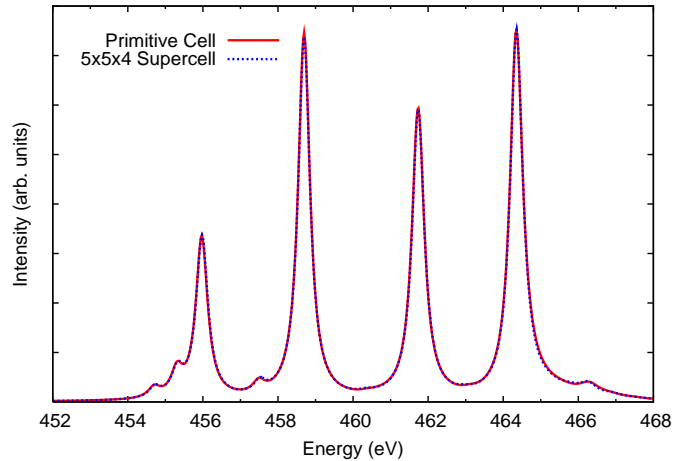}
\caption{{\bf Ti L-edge XAS of SrTiO$_{3}$.} Comparison of the Ti L-edge XAS of SrTiO$_{3}$ calculated
           from the primitive cell using {\sc abinit} and the serial version of {\sc ocean} and from the $5\!\times\!5\!\times\!4$ supercell using the OBFs and parallelized {\sc ocean} code.}
 \label{Ti_L-edge}
\end{figure}

Figure \ref{Ti_L-edge} presents the Ti L edge of pure SrTiO$_{3}$ calculated for both the 
primitive cell (one formula unit) using our previous version of {\sc ocean} \cite{vinson.ocean} 
and for the $5\times5\times4$ supercell (100 formula units) obtained with the code 
improvements described herein.  This comparison serves to verify that the fidelity of 
the computational scheme has been preserved through the modifications we have made and 
demonstrates the feasibility of calculating spectra of much larger systems than previously 
possible.  Doping SrTiO$_{3}$ yields a wide range of interesting physical properties 
and we imagine that in future investigations it will be fruitful to apply {\sc ocean} to studies 
of such systems.  Compounds of SrTiO$_{3}$ doped with transition metals are investigated for use 
in photocatalysis and as permeable membranes, among other uses.  Further, there is some evidence 
that doping with Mn, Fe and Co may yield dilute magnetic semiconductors \cite{norton.mn:sto, 
norton.co:sto, costa-pereira.thesis}.  The task of improving the performance of such materials is 
assisted by a better understanding of how the dopant element interacts with the host system.
It is now possible to model such spectra with a realistic, first-principles approach.

\begin{figure*}
  \includegraphics[angle=270,width=6.0cm]{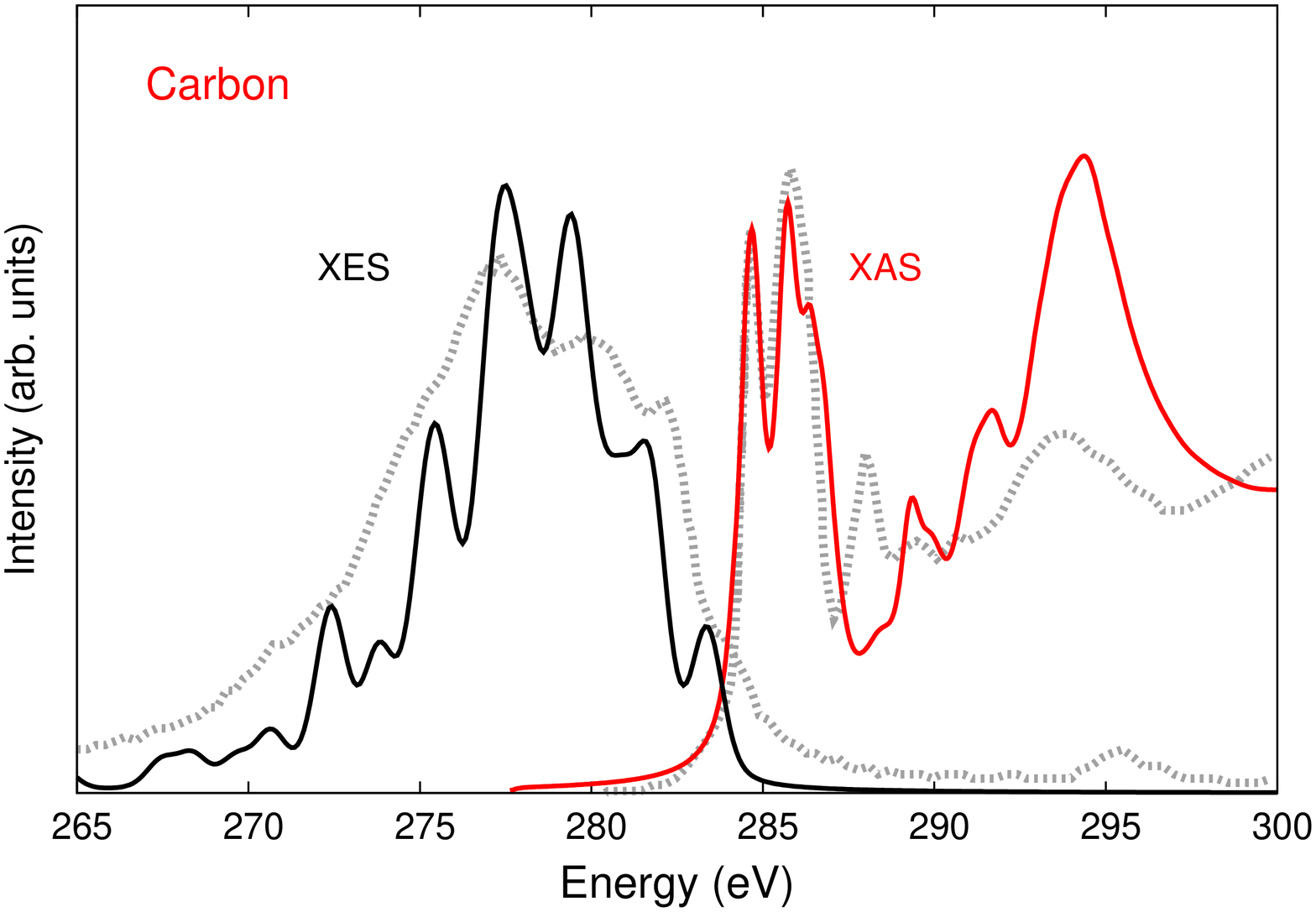}
  \includegraphics[angle=270,width=6.0cm]{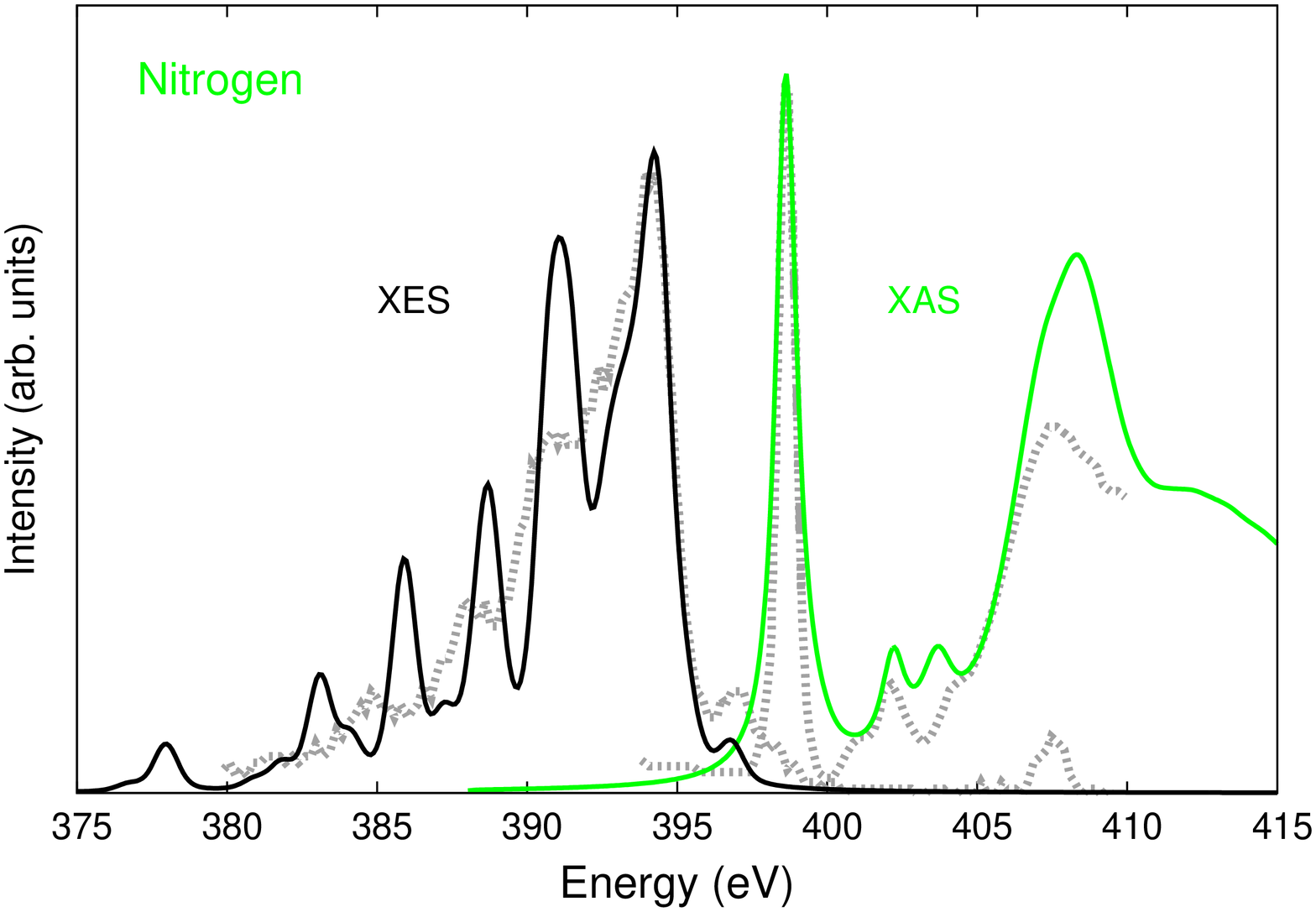}
  \includegraphics[angle=270,width=6.0cm]{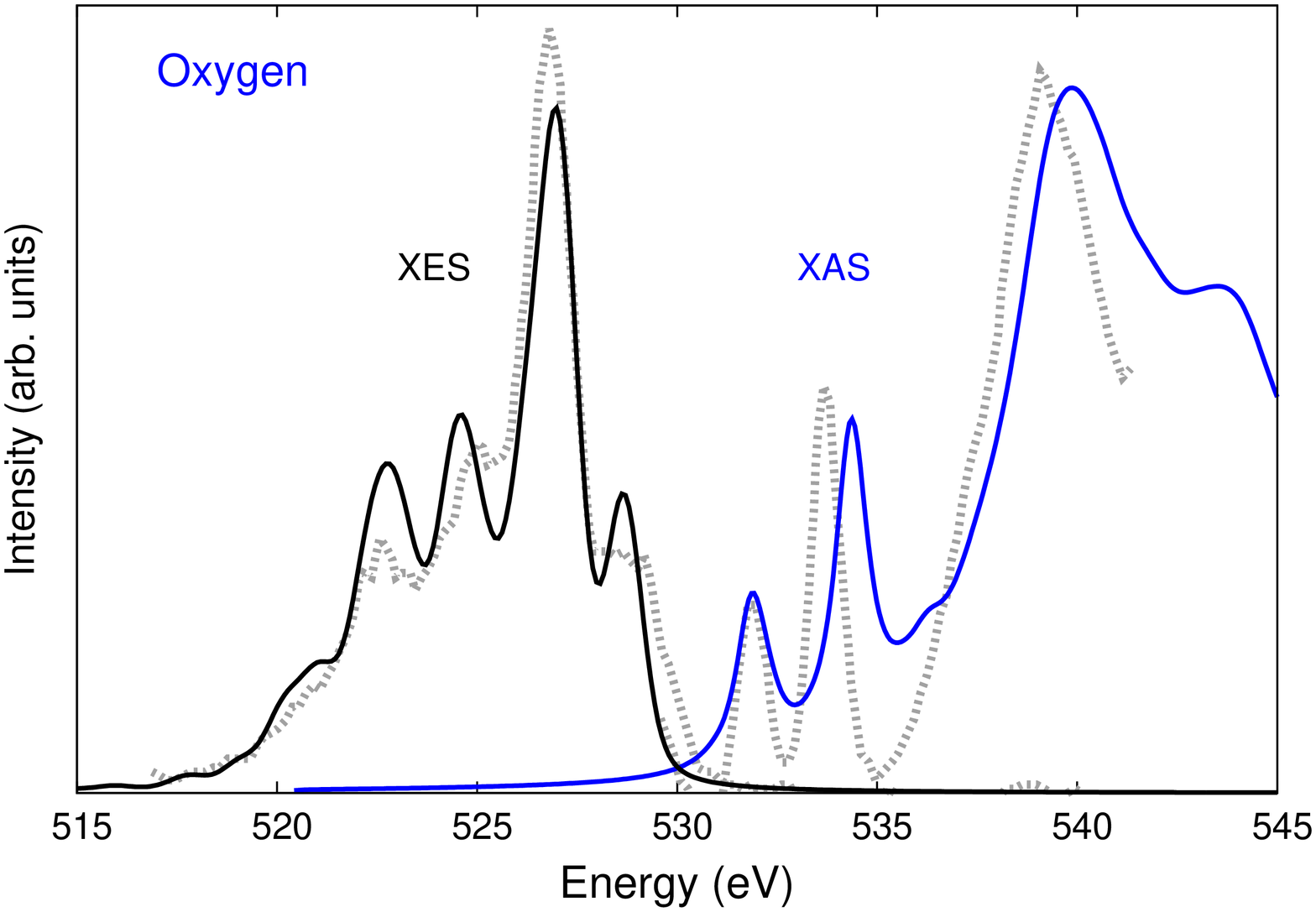}
  \caption{{\bf XAS and XES of Alq$_3$.}  K-edge XAS (color) and XES (black)
           of carbon (left, red), nitrogen (middle, green) and oxygen (right, blue) for gas-phase
           meridional Tris-(8-hydroxyquinoline)aluminum (Alq$_3$).  Grey curves show the experimental 
           data reproduced from reference \cite{demasi.alq3}.}
 \label{alq3}
\end{figure*}

As a second example system we consider the organic molecule Tris-(8-hydroxyquinoline)aluminum (Alq$_3$).  
Alq$_3$ has remained a leading molecule for the electron
transport and emitting layer in organic light emitting devices since it was first proposed
for this purpose \cite{TangVanSlyke}.  Despite numerous academic and industrial investigations
of such systems, significant problems remain, particularly regarding device lifetime.  The
Alq$_3$ molecule is susceptible to decomposition through reaction with water molecules 
\cite{alq3.hydrolysis} or metal atoms from the cathode layer \cite{shen.mg-alq3, car.mg-alq3}.  
X-ray spectroscopy is commonly used to further reveal the
interaction of water or metal ions with the Alq$_3$ molecule and the pathway by which
the complex decomposes.  However, results are difficult to properly interpret because such
experimental work is rarely coupled with calculated spectra and because the molecule is sensitive
to decomposition due to x-ray beam exposure.  In this section we demonstrate the ability of
{\sc ocean} to produce reliable XAS and XES spectra for this system.

Alq$_3$ exhibits two isomers, commonly referred to as facial and meridional.  The meridional isomer
is favored energetically and we consider only this structure.  In figure \ref{alq3} we present 
the C, N and O K-edge XAS and XES from the meridional Alq$_3$ molecule, which consists of 52 atoms.  
We treat the molecule in the gas-phase, though devices typically contain amorphous films of the molecule.
Since thermal atomic motion can noticeably impact spectral features for lighter elements we sum 
spectra from a series of configurations generated by a molecular dynamics (MD) simulation.  The 
spectra presented in figure \ref{alq3} show the average produced from 10 configurations of a MD 
simulation performed at 300 K within Quantum{\sc espresso}.  In
addition to averaging over the MD configurations, each spectrum is an average over each atomic
site in the molecule for the given element.  Thus, these results represent a series of 30
calculations for the N edge, 30 calculations for the O edge, and 270 calculations for the C
edge (keeping in mind that the same ground state DFT calculation can be used for all edges of a
given MD configuration).  Despite the large sampling required to produce these spectra
the calculations are not particularly burdensome.  After averaging over all samples and sites, 
self-energy corrections to the electronic structure were incorporated within the multi-pole 
self-energy scheme of Kas {\em et al.~} \cite{kas.mpse}.  Finally, an {\em ad hoc} rigid energy 
shift was applied to each spectrum to align the absolute energies with the experimental values.

Our calculations are generally in good agreement with the measured spectra of Ref.~\cite{demasi.alq3}.  
The primary structure of the XAS is reproduced for each 
element with only a few minor differences.  For carbon, the feature near 289 eV in our calculation 
appears in the experiment around 288 eV while, for nitrogen, the second feature is about 0.5 eV too 
low in energy in the calculation.  The three peaks of the oxygen XAS match the experimental spectrum 
quite closely.  The minor differences could originate in differences in electronic structure between 
the gas-phase, as we consider, and the condensed-phase material probed in experiment.  Additionally, 
we presently neglect vibronic coupling in the excited-state 
\cite{gelmukhanov.1977,tinte.vibronic, gilmore.vibronic} 
which can be particularly important in molecules with light elements 
\cite{ljungberg.2011,vinson.vibronic}.  
Nevertheless, agreement with experiment is generally quite favorable.

\section{Conclusion}

We have implemented a series of improvements that allow core-level spectral calculations based on solving 
the Bethe-Salpeter equation to be performed for much larger systems than previously possible.  Whereas 
{\sc ocean} was previously limited to systems of a few tens of atoms, here we have reported calculations 
on systems as large as a 5x5x4 supercell of SrTiO$_{3}$, which consists of 500 atoms and 3200 valence 
electrons and would allow for the direct simulation of doping at the 1~\% level.  This particular calculation required only 12.5~hours on 128 cores.
This enhanced capability makes spectral calculations on amorphous and dilute systems feasible.

We reduced the cost of the non-self consistent field DFT calculation through a {\bf k}-space interpolation scheme 
based on a reduced set of optimal basis functions.  This yielded a speed-up by 
a factor of 2-2.5 for large systems.  Parallelization of the screening calculation and evaluation of the BSE 
Hamiltonian provided further savings.  We find that the parallelization of the screening response scales 
only to a few dozen processors.  However, since the evaluation of the screening at this level of 
parallelzation has a limited cost compared to the initial DFT calculation this is not a significant 
concern at present. The BSE section of the code scales well to a few hundred processors with only a small growth in overhead that appears linear in processor count. When both MPI and OpenMP parallelization is used we achieved a speedup of 114x on 192 processors. However, when only MPI is used there is evidence that the overhead is growing superlinearly. There remains room for future improvement to reduce the MPI overhead. 

In addition to x-ray absorption spectra and x-ray emission spectra, inelastic x-ray scattering spectra 
may also be calculated with {\sc ocean}.  We view {\sc ocean} as uniquely capable of evaluating such 
spectra, particularly at L edges, with {\em ab initio} accuracy and predictive ability for complex and 
dynamic systems.  We envision use of {\sc ocean} to interpret data collected on a wide range of systems 
including {\em in operando} studies of fuel cell materials, photocatalysts, gas sensor and energy storage 
materials, as well as from liquid environments.

One should keep in mind that the Bethe-Salpeter equation is one of several approaches to calculating 
x-ray spectra.  While the BSE method holds advantages in being a predictive, first-principles approach, 
its two-particle formulation, in certain cases, is still a crude approximation to the actual many-body 
problem.  Contrary to this, many-particle approaches such as multiplet calculations and cluster models 
capture many-body physics more completely, though typically at the cost of band-structure effects.  The 
challenge for these local methods is to incorporate the electronic structure of the extended system in 
a scalable fashion.  It appears possible to make considerable progress toward this goal by working within 
a localized basis constructed from an extended electronic structure \cite{haverkort}.  For the 
BSE technique, it will be necessary to incorporate additional many-particle effects.  We expect that a better 
description of self-energy effects being made accessible by recent cumulant expansion development will 
prove advantageous in this respect \cite{kas.cumulants1, kas.cumulants2}.

The {\sc ocean} source code is now available for general use; for details and documentation see \cite{ObtainOCEAN}.

\section{Acknowledgments} This work is supported in part by DOE Grant DE-FG03-97ER45623 (KG, JJK, FDV, JJR).  
KG was additionally supported by the National Natural Science Foundation of China (Grant 11375127), 
the Natural Science Foundation of Jiangsu Provence (Grant BK20130280) and the Chinese 1000 Talents Plan.
Calculations were conducted with computing resources at Lawrence Berkeley National Laboratory and the 
National Energy Research Scientific Computing Center (NERSC), a DOE Office of Science User Facility 
supported by the Office of Science of the U.S. Department of Energy under Contract No. DE-AC02-05CH11231.

Certain software packages are identified in this paper to foster understanding. Such identification does not imply recommendation or endorsement by the National Institute of Standards and Technology, nor does it imply that these are necessarily the best available for the purpose. 

\vspace{0.25cm}

\bibliographystyle{elsarticle-num}
%\bibliography{bibfile}

\begin{thebibliography}{10}
\bibitem{core-hole}
When periodic boundary conditions are used spurious interactions between the
  exciton and its periodic images exist. In this case, it is necessary to use
  supercells to minimize these interactions.
\bibitem{diskio}
This is an effective bandwidth around 60~MB/s which compares poorly to
  commercial spinning disk contiguous read speeds which range 200-300~MB/s.
\end{thebibliography}

\end{document}